\documentclass[12pt]{article}
\usepackage{newtxmath}
\usepackage{mathbbol}
\usepackage{bbm}
\usepackage{booktabs}
\usepackage{amsmath}
\usepackage{booktabs}
\usepackage{geometry}
\usepackage{fancyhdr}
\usepackage{graphicx}
\usepackage{subfigure}
\usepackage{float}
\usepackage{color}
\usepackage[colorlinks,bookmarksopen,bookmarksnumbered,citecolor=blue, linkcolor=red, urlcolor=blue]{hyperref}

\newcommand\keywords[1]{\textbf{Keywords}: #1}

\title{Reconstruction of gene regulatory network via sparse optimization}

\date{}

\begin{document}

\maketitle

\begin{center}
    {\large\lineskip .5em Jiashu Lou\textsuperscript{1},~Leyi Cui\textsuperscript{1},~Wenxuan Qiu\textsuperscript{1}} \\
    
\textsuperscript{1}~{College of Mathematics and Statistics, Shenzhen University} \\
\end{center}

\begin{abstract}
   In this paper, we tested several sparse optimization algorithms based on the public dataset of the DREAM5 Gene Regulatory Network Inference Challenge. And we find that introducing 20\% of the regulatory network as a priori known data can provide a basis for parameter selection of inference algorithms, thus improving prediction efficiency and accuracy. In addition to testing common sparse optimization methods, we also developed voting algorithms by bagging them. Experiments on the DREAM5 dataset show that the sparse optimization-based inference of the moderation relation works well, achieving better results than the official DREAM5 results on three datasets. However, the performance of traditional independent algorithms varies greatly in the face of different datasets, while our voting algorithm achieves the best results on three of the four datasets.\par

To explore the adaptability of different algorithms for different size networks, we conducted experiments based on simulated data and found that the coordinate descent-based Lasso algorithm and OMP algorithm are better for small-scale network reasoning; while the improved algorithms such as SP and CosaMP have higher accuracy for large-scale networks. Meanwhile, we focus on the noise in the gene regulation process, and we find that in a low-noise situation, algorithms such as OMP with good theoretical convergence support have higher prediction accuracy. And for the realistic gene regulation with high noise, using voting algorithm can give a good result. The voting algorithm always gives a better prediction result and has good accuracy variance and stability for repeating the experiment many times.\par

Finally, we make the code used in the experiments into a python package and open source it on Pypi and Github.
\end{abstract}
\newpage
\tableofcontents
\newpage

\section{Introduction}
\subsection{Gene regulation network}
For certain diseases or traits, genes play a crucial role in their expression. And the interactions between genes, i.e., gene regulatory networks (GRNs), have become a recent research hotspot. With the availability of high-throughput gene expression data and the substantial increase in arithmetic power, it is possible to reconstruct large-scale gene regulatory networks\cite{ref1}. Due to the nature of gene expression data, providing information about the abundance of mRNAs only rather than binding information, gene regulatory networks defined in the above sense provide information about regulatory interactions between regulators and their potential targets; gene-gene interactions, and potential protein-protein interactions. In this paper, we refer to the network inferred in this way as a \textit{gene regulatory network}\cite{ref32}.\par

In gene regulatory networks, genes can be divided into two categories. Transcription factors (TF), also known as trans-acting factors, are DNA-binding proteins that specifically interact with the cis-acting elements of eukaryotic genes and have an activating or inhibiting effect on gene transcription. The gene that receives this activation or repression is referred to as the target gene. It is important to note that transcription factors themselves may also be target genes, i.e., there may be mutual regulation in the regulatory network.\par

\subsection{Mathematical Models}
In a biological system consisting of $n$ transcription factors and $m$ target genes, there are at most $m*n$ groups of regulatory relationships. However, the fact is that the number of groups of regulatory relationships will be much smaller than $m*n$, i.e. such a regulatory network is sparse. Current research is mainly based on differential equations\cite{ref11}\cite{ref12}, Boolean networks\cite{ref13}, mutual information\cite{ref16}, and Bayesian networks\cite{ref14}\cite{ref15}. However, these methods do not make good use of the sparse nature of gene regulatory networks, resulting in unsatisfactory accuracy or excessive time consumption.\par

We assume that the regulation between genes is linear. 
However, If we have 1000 target genes, this means that we need to make 1,000,000 independent measurements. Even though technological advances in measuring genome-wide expression, such as DNA microarray technology, have facilitated simultaneous expression measurements of a large number of genes, this still requires significant resources\cite{ref29}. Therefore, in a realistic situation, we will get a much smaller number of experimental samples than the number of target genes, which means the equation(\ref{equ1}) 

\begin{equation}\label{equ1}
    Ax=b+\varepsilon
\end{equation}

is a ill-posed problem. The mathematical description of the problem is as equation(\ref{equ2})

\begin{equation}
A \in \mathbb{R}^{m \times n}, \quad x \in \mathbb{R}^{n}, \quad b \in \mathbb{R}^{m} \label{equ2}
\end{equation}

Since $rank(A)<n$, the equation $Ax = b+\varepsilon $ has an infinite set of solutions. At this point, if when $x$ is sparse, the equation has a unique solution. We call this class of problems as compression-aware or sparse reconstruction problems.\par

\begin{figure}[H]
    \centering
\begin{center}
    \includegraphics[scale = 0.2]{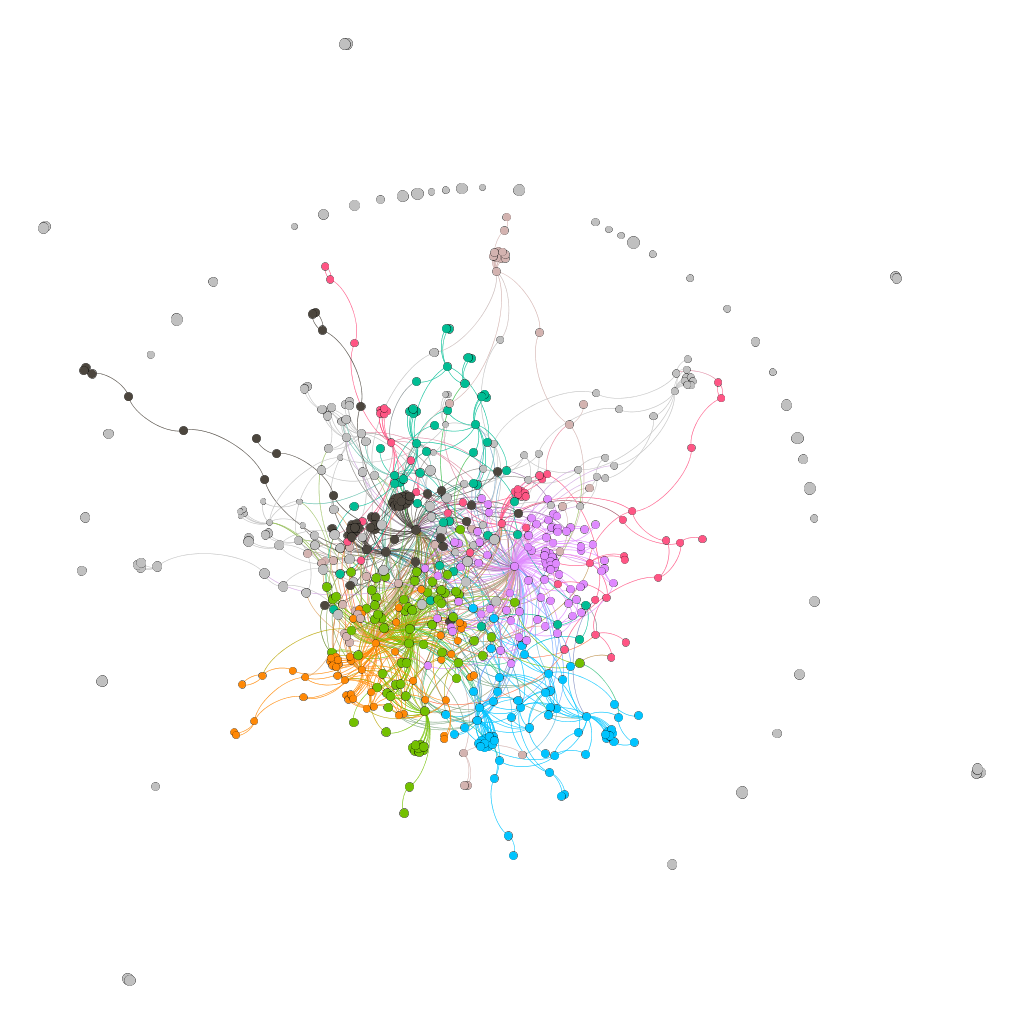}
\end{center}
    \caption{\textbf{Gene regulatory relationships:} Playing the role of $x$ in the above set of linear equations, the colored dots are the factors involved in the regulatory relationship and the gray dots represent the factors that are not involved in the regulation. From the figure, we can find that this regulatory relationship is sparse.}
\end{figure}

\subsection{This paper}
In this paper, we will use the DREAM5\cite{ref3}\cite{ref26} challenge dataset as experimental data. Conceptually, We consider GNs as a directed and unprivileged network\cite{ref30}. Both this paper and DREAM5 challenge focused on the existence of regulatory relationships rather than the direction and intensity of regulation.\par

First, we designed the workflow for reasoning about the regulatory network from the expression matrix and tested various sparse optimization algorithms. We then bagged these algorithms using voting method to achieve better results for any dataset. In these works, determining the sparsity is a difficult task. Therefore, we assume that 20\% of the gene regulatory network is known, and determine the optimal sparsity by introducing known prior data.\par
Our voting algorithm performs well on the DREAM5 dataset. Next, we want to explore the effect of the characteristics of the dataset on the accuracy of the different algorithms.\par
Due to the large cost of real data acquisition, we then synthesized simulated data based on linear stochastic synthesis algorithms and varied their sparsity, the number of target genes, sample size, and other factors to explore the factors influencing the accuracy.\par
Finally, we integrated the above algorithms into an easy-to-use package and open-sourced it.

\section{Materials \& Methods}
\subsection{Data sets from DREAM5}
The data is a publicly available dataset from the Dream Challenge, with four sets of data. The first of these sets is computer-simulated data. The last three data sets are derived from microbial and real pathogen sequencing results, and the specific data sources are shown in the following table.

\begin{table}[H]
\begin{center}
\begin{tabular}{@{}lllll@{}}
\toprule
\textbf{Network names}         & \textbf{Network1} & \textbf{Network2} & \textbf{Network3} & \textbf{Network4} \\ \midrule
\textbf{source} & Computer simulations   & Staphylococcus aureus  & Escherichia coli     & Brewer's yeast     \\ \bottomrule
\end{tabular}
\caption{Data exploration}
\end{center}
\end{table}\par
For the golden standard regulatory network, the gene regulation data of E. coli and Saccharomyces cerevisiae are experimentally validated\cite{ref4}\cite{ref5}. In contrast, for computer simulated data, the network is known. The S. aureus network cannot be measured accurately because there are relatively few experimentally supported interactions available. Nevertheless, DREAM confirmed the accuracy of the predictions by integrating the predictions of all teams using an average ranking approach and by evaluating the network using a large number of computationally derived interactions from the RegPrecise\cite{ref6} database. Since the data are derived from real experiments, the size and sparsity of the data are very irregular, which are as follows.

\begin{table}[H]
\begin{center}

\begin{tabular}{@{}lllll@{}}
\toprule
\textbf{Network names}      & \textbf{Network1} & \textbf{Network2} & \textbf{Network3} & \textbf{Network4} \\ \midrule
\textbf{TF numbers}  & 195               & 99                & 334               & 333               \\
\textbf{Gene numbers}   & 1643              & 2810              & 4511              & 5950              \\
\textbf{Sample numbers}   & 805              & 161              & 805              & 536              \\
\textbf{Regulating network sparsity} & 0.01252            & 0.001581          & 0.001371          & 0.001927           \\ \bottomrule
\end{tabular}
\caption{Network sharp information}
\end{center}
\end{table}\par
 The sparsity of all four networks is between 0.001 and 0.002. In the next experiments, we will first make predictions for these four modulated networks and compare them with the public rankings of the DREAM community, and then we will try to generate the data by ourselves using linear synthesis and explore the relationship between accuracy and factors such as data size and sparsity.

\subsection{Scoring Metrics}
Since the reference network matrix in this dataset is a matrix consisting of only 0 and 1, where 1 represents the presence of a relationship and 0 represents the absence of a relationship. That is, our prediction target is an undirected and unweighted network. For this matrix, we use the AUC (area under the ROC curve) to measure the accuracy of the prediction, which is often used to evaluate the accuracy of classification models, and the problem can also be regarded as classifying the relationships between genes into "conductive" and "non-conductive" categories. The AUC can also be used to evaluate the relationship between genes as "conductive" and "not conductive". The calculation formula is as follows.

\begin{equation}
    AUC=\frac{\sum  { pred }_{ {pos}}> { pred}_{ {neg }}}{{positiveNum * negativeNum }}
\end{equation}\par

Notice that in this problem positive and negative cases are uneven (due to the sparsity of the network), while the AUC calculation method considers the classification ability for both positive and negative cases, and is still able to make a reasonable evaluation of the classifier in the case of unbalanced samples. Therefore, AUC is not sensitive to whether the sample categories are balanced or not, and is suitable as a scoring criterion for this task. Furthermore, real biological data and simulated data can, and should, be used for the assessment of networks. For real biological data this allows to assess the biological relevance of inferred networks\cite{ref1}. However, due to the lack of data, we are not going to test reasoning result in a realistic biological systems.

\subsection{Sparse optimization model}
This class of algorithms always accepts an $m*n$ matrix $\mathbf{A}$ with a $1*m$ vector $\mathbf{b}$ and outputs a sparse $1*n$ dimensional vector $\mathbf{x}$. In this problem, the non-zero elements in the vector $\mathbf{x}$ represent the presence of a regulatory relationship between the transcription factors in the matrix $\mathbf{A}$ and the target genes in $\mathbf{b}$. And based on how to select out the non-zero elements in $\mathbf{x}$ the algorithm can be divided into two categories: greedy algorithms based on matching tracking and compressed perception algorithms based on convex optimization. 
These two methods are described separately in the following. To distinguish our voting algorithm introduced later, we call the sparse optimization algorithms in this section independent algorithms.
\subsubsection{Least squares optimization}
The essence of the algorithm based on convex optimization is to solve the regularized regression problem with $L1$ parametrization showed as equation (\ref{equ3}):

\begin{equation}
    \min _{w} \frac{1}{2 n_{{samples }}}\|X w-y\|_{2}^{2}+\alpha\|w\|_{1} \label{equ3}
\end{equation}

Since this loss function is a non-convex function, we need to find an optimization algorithm to solve it. Commonly used algorithms are coordinate descent method, ISTA algorithm, etc.\par
Coordinate descent is a non-gradient optimization algorithm that performs a one-dimensional search along a coordinate direction at the current point in each iteration to find the local minima of a function. Different coordinate directions are recycled throughout the process.It has been proved by Jerome Friedman(2010)\cite{ref7} that the coordinate descent method is able to converge to a sparse solution in solving the Lasso problem.\par
The ISTA algorithm continues to solve the least squares problem with L1 parametrization with the iterative formulation:

\begin{equation}
    \mathbf{x}_{k+1}=\operatorname{soft}_{\lambda t}\left(\mathbf{x}_{k}-2 t \mathbf{A}^{\mathrm{T}}\left(\mathbf{A} \mathbf{x}_{k}-\mathbf{y}\right)\right)\label{equ4}
\end{equation}

where $soft$ is the soft threshold operator function:

\begin{equation}
    soft_{T}(\mathbf{x})=sign\left(x_{i}\right)\left(\left|x_{i}\right|-T\right)\label{equ5}
\end{equation}

Beck(2009) had proved that when the iteration step is taken as the inverse of the Lipshitz constant\cite{ref8}, the time complexity of the algorithm is $O(\frac{1}{k})$.

\subsubsection{greedy algorithms}
The core idea of the greedy algorithm is to select the atom that is closest to the residual of the current sample in each iteration. The resulting algorithms are OMP, CoSaMP, SP, IHT, etc.The core idea of this type of algorithm is to select one or more atoms with the highest correlation in each iteration to be added to the solution set as non-zero variables. \par
Take OMP as an example. For matrix A and residuals r, first: 

\begin{equation}
    i=\arg \min_{i}\left|A_{i}^{T} r_{k}\right|\label{equ6}
\end{equation}

Then:
\begin{equation}
    P_{k}=A_{S}\left(A_{S}^{T} A_{S}\right)^{-1} A_{S}^{T}, \quad r_{k}=\left(I-P_{k}\right) b\label{equ7}
\end{equation}

After repeating the above steps k times, return:

\begin{equation}
    x_{S}=\left(A_{S}^{T} A_{S}\right)^{-1} A_{S}^{T} b\label{equ8}
\end{equation}

Now $x$ contains the $k$ most relevant elements. Joel A Tropp(2004) had proved OMP algorithm can converge in the absence of noise\cite{ref9}. T. T. Cai(2011) had proved the convergence in the presence of noise\cite{ref10}.\par
The CosaMP algorithm, on the other hand, modifies the approach of selecting elements based on OMP. It calculates the subscript of $\theta$ by the equation(\ref{equ9}) and selects the first $K$ larger as the final output solution.

\begin{equation}
    \hat{\theta}_{t}=\arg \min _{\theta_{i}}\left\|y-A_{t} \theta_{t}\right\|=\left(A_{t}^{T} A_{t}\right)^{-1} A_{t}^{T} y\label{equ9}
\end{equation}

Needell, D(2009) pointed out that CoSaMP delivers the same guarantees as the best optimization-based approaches. Moreover, this algorithm offers rigorous bounds on computational cost and storage. It is likely to be extremely efficient for practical problems because it requires only matrix-vector multiplies with the sampling matrix\cite{ref17}.\par
The SP algorithm is very similar to the CosaMP algorithm, and their most important differences are, in each iteration, in the SP algorithm, only $K$ new candidates are added, while the CoSAMP algorithm adds $2K$ vectors.Dai W(2009) recognized that the SP algorithm has similar properties and advantages and disadvantages to the CoSaMP algorithm, but the SP algorithm is computationally more efficient\cite{ref18}.\par
IHT is not a convex optimization algorithm, it is similar to OMP, which is an iterative algorithm, but it is a greedy algorithm derived from an optimization problem. Therefore, it has the characteristics of two types of algorithms.IHT algorithm for solving least squares problems with L1 parametrization:

\begin{equation}
    C_{\ell_{0}}(\mathbf{y})=\|\mathbf{x}-\Phi \mathbf{y}\|_{2}^{2}+\lambda\|\mathbf{y}\|_{0}\label{equ10}
\end{equation}

This is not an easy problem to optimize, so the following equation is introduced as an alternative objective function.

\begin{equation}
    C_{\ell_{0}}^{S}(\mathbf{y}, \mathbf{z})=\|\mathbf{x}-\Phi \mathbf{y}\|_{2}^{2}+\lambda\|\mathbf{y}\|_{0}-\|\Phi \mathbf{y}-\Phi \mathbf{z}\|_{2}^{2}+\|\mathbf{y}-\mathbf{z}\|_{2}^{2}\label{equ11}
\end{equation}

It is easy to prove\cite{ref19} the equations(\ref{equ12}) and (\ref{equ13}):

\begin{equation}
    C_{\ell_{0}}^{S} \ge C_{\ell_{0}} \quad \forall z \label{equ12}
\end{equation}

\begin{equation}
    \exists z = y \quad s.t. \quad C_{\ell_{0}}^{S} = C_{\ell_{0}} \label{equ13}
\end{equation}

According to the Majorization-Minimization\cite{ref20} idea, we can optimize $C_{\ell_{0}}^{S}$ as an alternative function. The final obtained iterative formula is as follows.

\begin{equation}
    \mathbf{y}^{n+1}=H_{M}\left(\mathbf{y}^{n}+\Phi^{H}\left(\mathbf{x}-\Phi \mathbf{y}^{n}\right)\right) \label{equ14}
\end{equation}

\begin{equation}
	H_{M}\left(y_{i}\right) = \begin{cases}
	      0, & if \left|y_{i}\right|<\lambda_{M}^{0.5}(\mathbf{y}) \\
	      y_{i}, & if \left|y_{i}\right| \geq \lambda_{M}^{0.5}(\mathbf{y})
		   \end{cases}
\end{equation}\label{equ15}\par

Since there is no evidence that a sparse optimization algorithm is the best one (and there is no algorithm that is optimal for any data according to the "no free lunch" theorem\cite{ref27}), we will test each algorithm in the following.

\subsubsection{Voting algorithm}
For the above methods, there is no study to prove which is the optimal method for gene regulatory network reconstruction. We also found that the optimal method varies for different data through subsequent experiments. Therefore, we will develop a voting algorithm based on the above methods to obtain better prediction results for any data set.\par
The idea of the voting algorithm is simple: since every algorithm has some possibility of making mistakes, the result that is approved by most algorithms should have relatively high accuracy. For the modulation relation at the $(i,j)$ position, we have the following algorithm:

\begin{equation}
	N_{i,j} = \begin{cases}
	      1, & if \sum_{k=1}^{n}N_{k,i,j}>a \\
	      0, & if \sum_{k=1}^{n}N_{k,i,j}\le a	
		   \end{cases}
\end{equation}\label{equ16}

The meaning of the parameter $a$ is that when $a$ of the algorithms participating in the voting believe that a moderation relation exists here, it is recognized as such. The setting of $a$ also affects the sparsity of the final result. A smaller value of $a$ yields a denser matrix - the result is more likely to contain the correct moderation relation but is also more likely to treat a non-existent moderation relation as the present.\par

Take a system containing five transcription factors as an example: Transcription factors A, B, C, D, and E may have potential regulatory relationships for the target gene T. Five algorithms are used to make predictions for this network. Assume that the gold standard for this set of regulatory relationships is the equation:

\begin{equation}
    A*\begin{bmatrix}
 0 &0  & 1 &1  &0
\end{bmatrix}^{T}=b 
\end{equation}\label{equ17}

The following predictions were given by the five algorithms, $S_1$ to $S_5$:

\begin{figure}[H]
    \centering
\begin{center}
    \includegraphics[scale = 0.1]{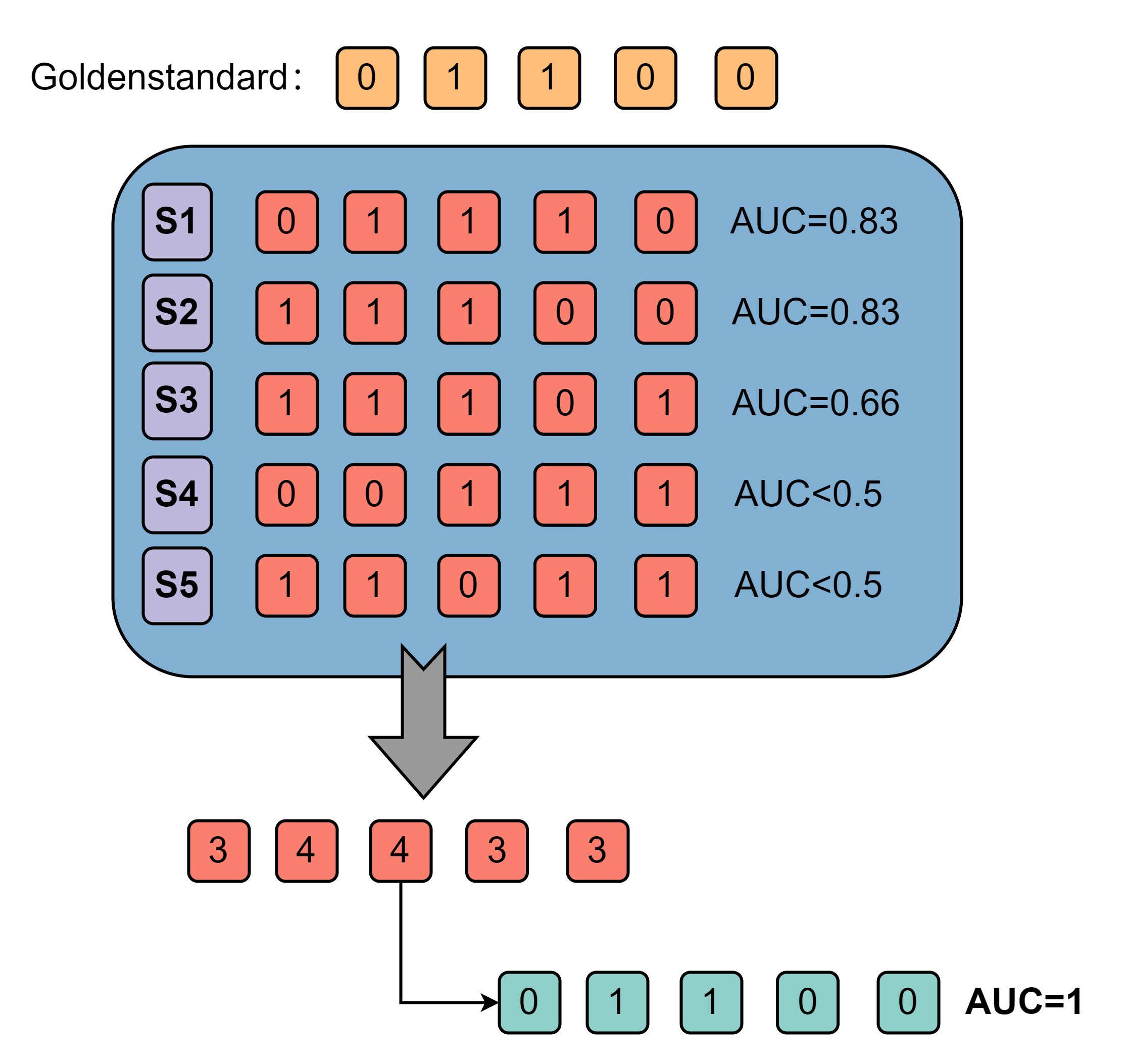}
\end{center}
    \caption{An example of voting algorithm}
\end{figure}\par

As we can see from the above figure, the algorithms from A to E contribute 5 weaker predictions respectively, which even have $AUC < 0.5$. But after the integration of the voting algorithms, the final result of the perfect classifier with $AUC=1$ is obtained.\par
The advantage of the voting algorithm is that it has a certain error tolerance for each independent algorithm, and since it is very unlikely that each independent algorithm will make an error at the same time (if so, it is more important to check the original data set or data processing for errors), then the voting algorithm will definitely output a better result. Voting algorithms are generally employed for classification tasks, such as Li, J (2019)\cite{ref28} used a voting algorithm for the classification task of thermophilic proteins. And as mentioned above, the prediction of gene regulatory networks has similarities with the classification task, so the voting algorithm will still be effective for gene regulatory networks based on experiments demonstrated below.

\subsection{work flow}
Solving this problem is equivalent to solving a system of linear equations(\ref{equ1}).
We want to obtain the sparse solution of $x$. where $A$ is the expression matrix of transcription factors, $x$ is the sparse gene regulation matrix, and $b$ is the expression data of the target gene.

\begin{figure}[H]
    \centering
\begin{center}
    \includegraphics[scale = 0.1]{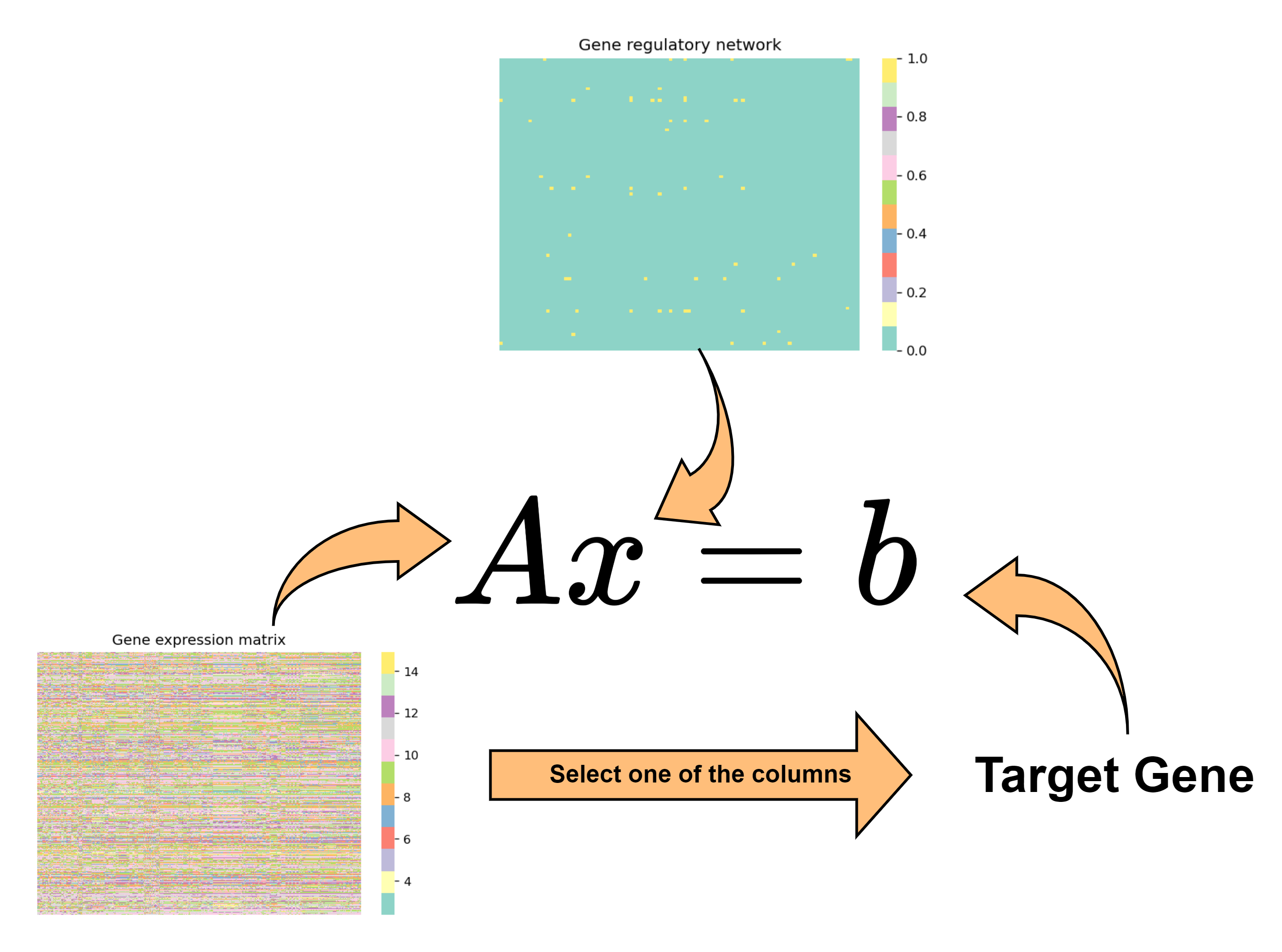}
\end{center}
    \caption{\textbf{Linear equations:}This image gives the general form of the gene regulation network prediction problem and what each variable in this system of linear equations represents}
\end{figure}\par

We will normalize the entire network inference process to facilitate the subsequent packaging of the code into packages for distribution. First, we will receive four files, which are the gene expression matrix, the transcription factor list, the gene list, and the gold standard for gene regulatory networks. The gold standard of the gene regulatory network can be missing, but then it will not score the final prediction results. The rows and columns contained in each file are shown in the following table:

\begin{table}[H]
\begin{center}

\begin{tabular}{@{}cc@{}}
\toprule
\textbf{File names}        & \textbf{Introduction}                                                                                                                                                                                                                                                                                                                                                          \\ \midrule
\textbf{Gene names}        & A column of genetic names.\\                                                                                                                                                                                                                                                                                                                                                     \\
\textbf{TF names}          & A column of transcription factors names.\\                                                                                                                                                                                                                                                                                                                                       \\
\textbf{Expression matrix} & \begin{tabular}[c]{@{}c@{}}Each column represents a gene and each row represents \\ the expression level of different genes in the same sample.\\ \end{tabular}  \\                                                                                                                                                                                                                 \\
\textbf{network}           & \begin{tabular}[c]{@{}c@{}}A side list, the first column represents the transcription factor,\\  the second column represents the target gene \\ regulated by that transcription factor, \\ and the third column is the type of regulation: \\ 1 indicates the presence of a regulatory relationship \\ and 0 indicates the absence of a regulatory relationship.\\\end{tabular} \\ \bottomrule
\end{tabular}
\caption{Meaning of the input files}
\end{center}
\end{table}\par

After importing the data, we first process the network data (if available) into the form of a matrix. Where the (i,j) position if 1, indicates that the No.j transcription factor has a regulatory relationship with the No. i target gene. We then traversed the entire gene expression matrix, testing each gene as a target gene from the first to the last. It is worth noting that since the target gene itself may also be a transcription factor, in this case, we must first manually delete the column where this gene is located and then insert a zero at that position in the result matrix. otherwise, it will result in an unusually large coefficient at that point. After testing all the independent algorithms, we bagged them using the voting algorithm, calculated their AUC values separately, and performed a comparative analysis.

\begin{figure}[H]
    \centering
\begin{center}
    \includegraphics[scale = 0.1]{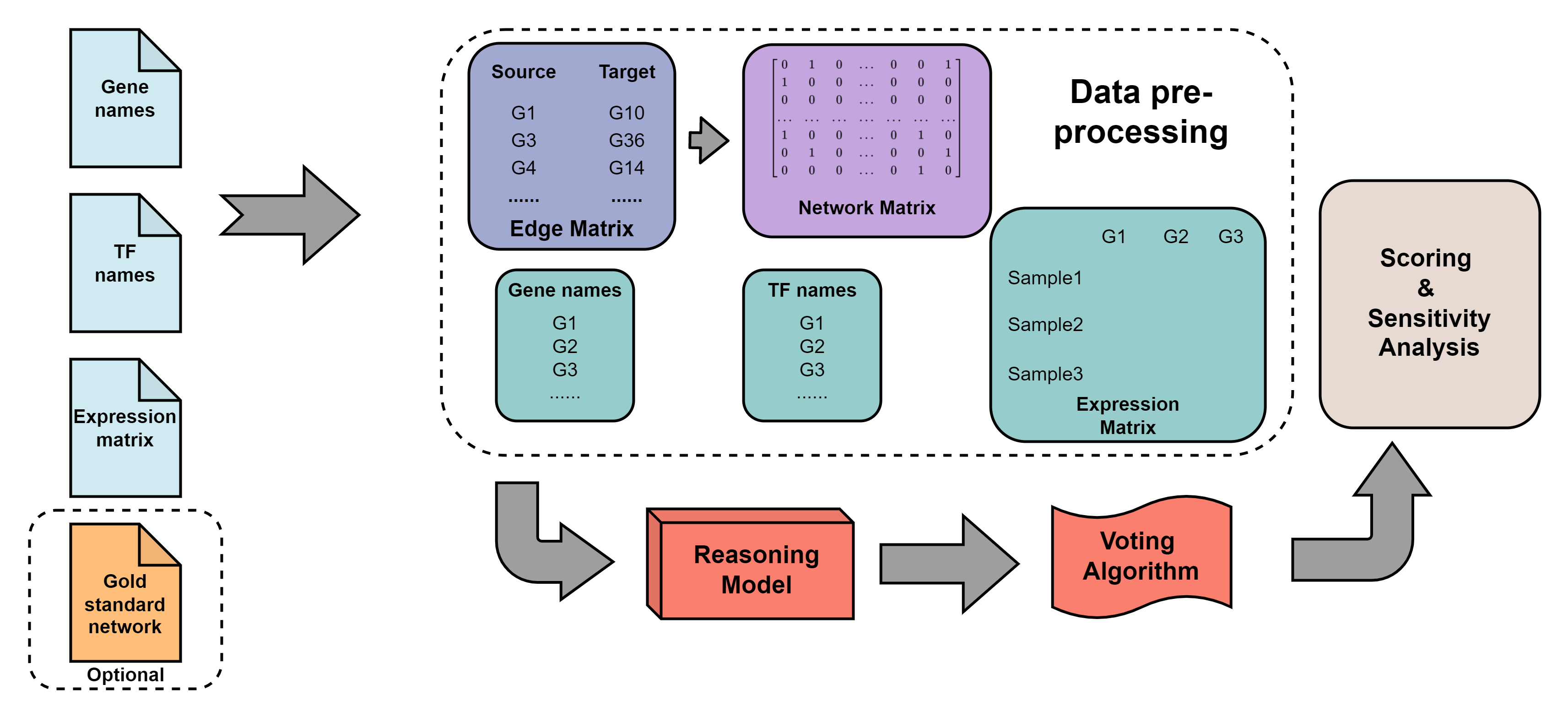}
\end{center}
    \caption{Our workflow}
\end{figure}\par

\subsection{Sparse tools box}
A commonly used gene regulatory network inference package is currently the R-based Genie3 package. In Genie3, a random forest was trained to predict gene regulatory network and putative transcription factors are selected as tree nodes if they consistently reduce the variance of the target\cite{ref31}. The Genie3 method was also used to participate in the official DREAM5 rankings and achieved the first overall ranking. However, since it is written based on R, it runs very slowly. Python, on the other hand, also being a popular programming language in the data science field, is able to provide several times the running speed of R. So, we will organize the main code used in this paper and make it into a Python package.
\subsubsection{Random data generation}
To explore the effects of network size, the number of transcription factors, and other factors on prediction accuracy. We synthesized simulated data based on the following principles. 
The task of the DREAM challenge is to solve for $x$ when $A$ and $b$ are known. to explore the effects of the shapes of both $A$ and $b$ and the sparsity of $x$ on the experimental results, we generate $A$ randomly and a sparse matrix $x$ with specified sparsity and $Ax+\mu$ as $b$, the relationships are shown below:
\begin{equation}
\begin{aligned}
& A*x = b+\mu \\
& \mu \sim N(normalLoc, normalScale) \\
\end{aligned}
\end{equation}

The algorithm generates 10 transcription factors, 200 samples, and 2000 target genes by default. The expression matrix obeys a [0,1] uniform distribution and the regulatory network obeys a [-1,1] uniform distribution. For each set of regulatory relationships, $20\%\pm 10\%$ of the transcription factors are involved in regulation, that is, the sparsity is around 10\% to 30\%.\par

\subsubsection{Reasoning algorithms}
In addition to the random data generation function, the gene regulatory network reconstruction algorithm mentioned above is also integrated into the package. We combine the above-mentioned independent algorithms with the voting algorithm, and the user only needs to simply call the API to implement the network inference process. The random data generation algorithm and the inference algorithm are also shown to be stable and efficient in the following section. \par
For detailed usage and the format of input and output can be found in https://pypi.org/project/Sparse-Optimization-Toolbox/

\section{Result}
\subsection{Parameter Selection}
In this problem, the most important hyperparameter is the sparsity of the matrix, and we need to tell the algorithm how many non-zero elements are contained in the final regulatory network. For different matrices, the sparsity varies and is not regular. However, based on biological experiments, some of the regulatory relationships in a system are already proven. Therefore, we take the top 20\% of the regulatory relationships in the data set as a priori data and calculate their AUC values by varying the sparsity, to choose a better sparsity for each network.The sparsity there refers to:

\begin{equation}
    sparsity = \frac{S_a}{S_b}
\end{equation}\label{sparsity}

Where $S_a$ is the number of non-zero items in a matrix and $S_b$ is the numbers of items in a matrix.
\par
By experimenting with different sparsity selected with 20\% of the data, we selected their optimal sparsity for the four networks. Since sparsity here denotes the proportion of non-zero elements in the matrix. We believe that the optimal sparsity should be the same for different algorithms. We use the SP algorithm as the test algorithm when selecting the parameters because the SP algorithm is less time consuming. The experimental results are as follows:
\begin{figure}[H]
\centering
\subfigure[Network1]{
\includegraphics[scale = 0.4]{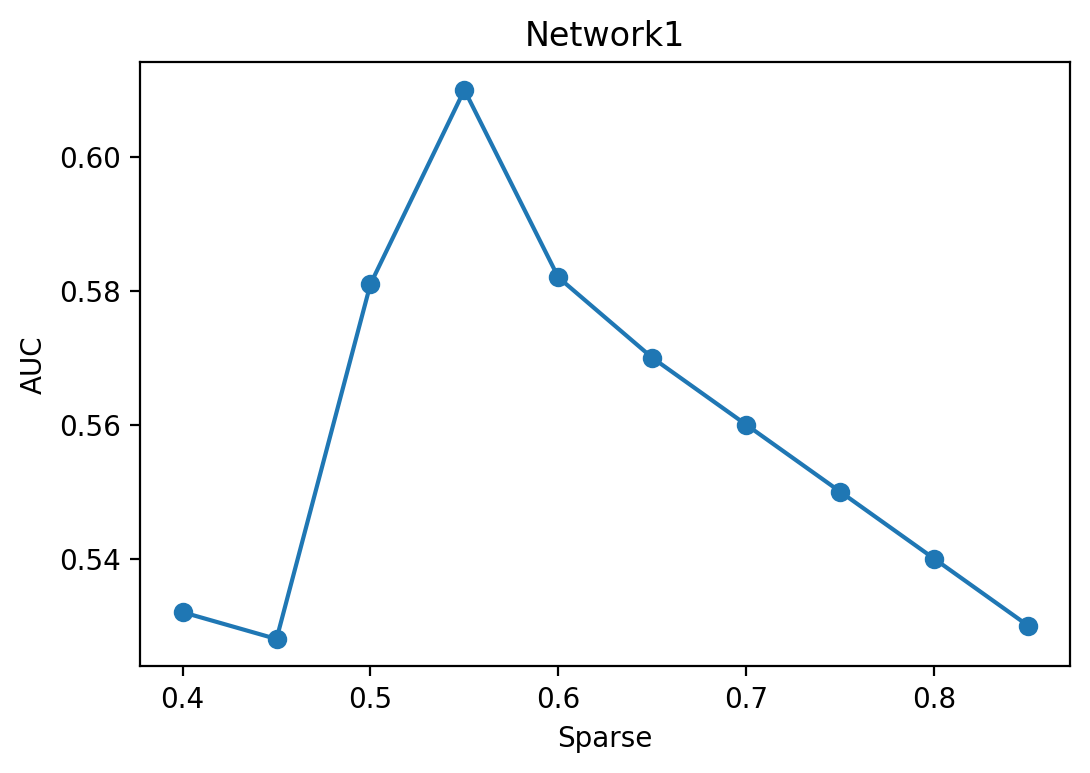}
}
\quad
\subfigure[Network2]{
\includegraphics[scale = 0.4]{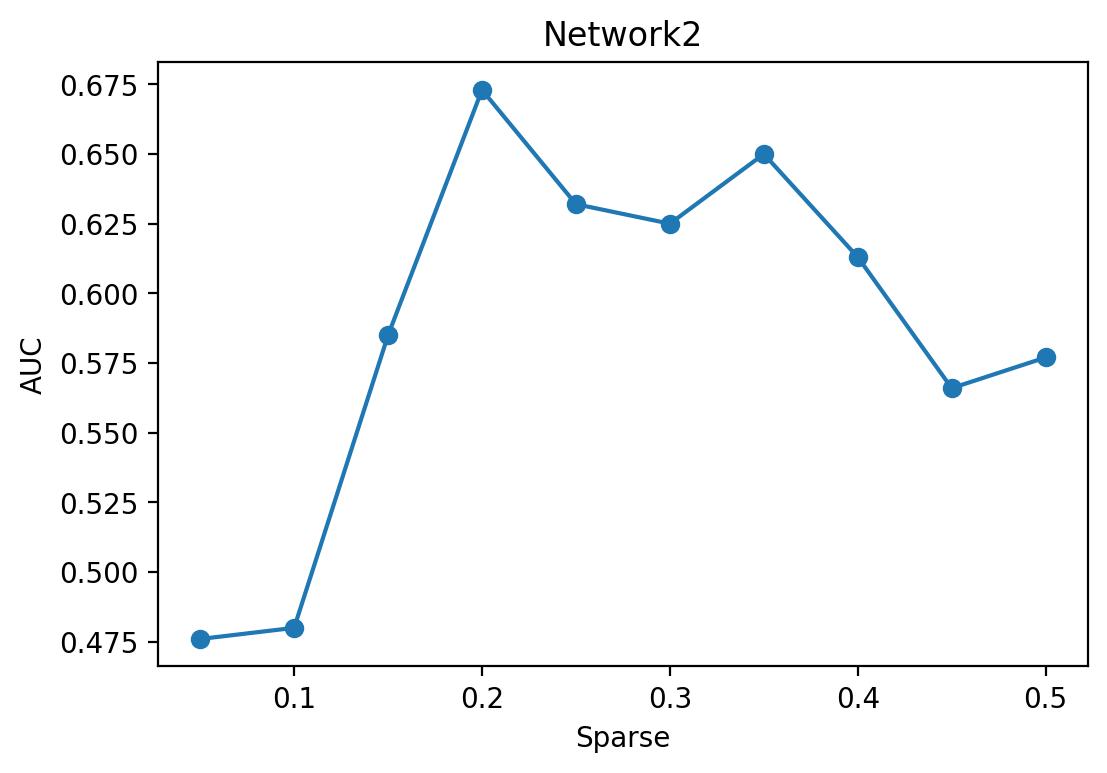}
}
\quad
\subfigure[Network3]{
\includegraphics[scale = 0.4]{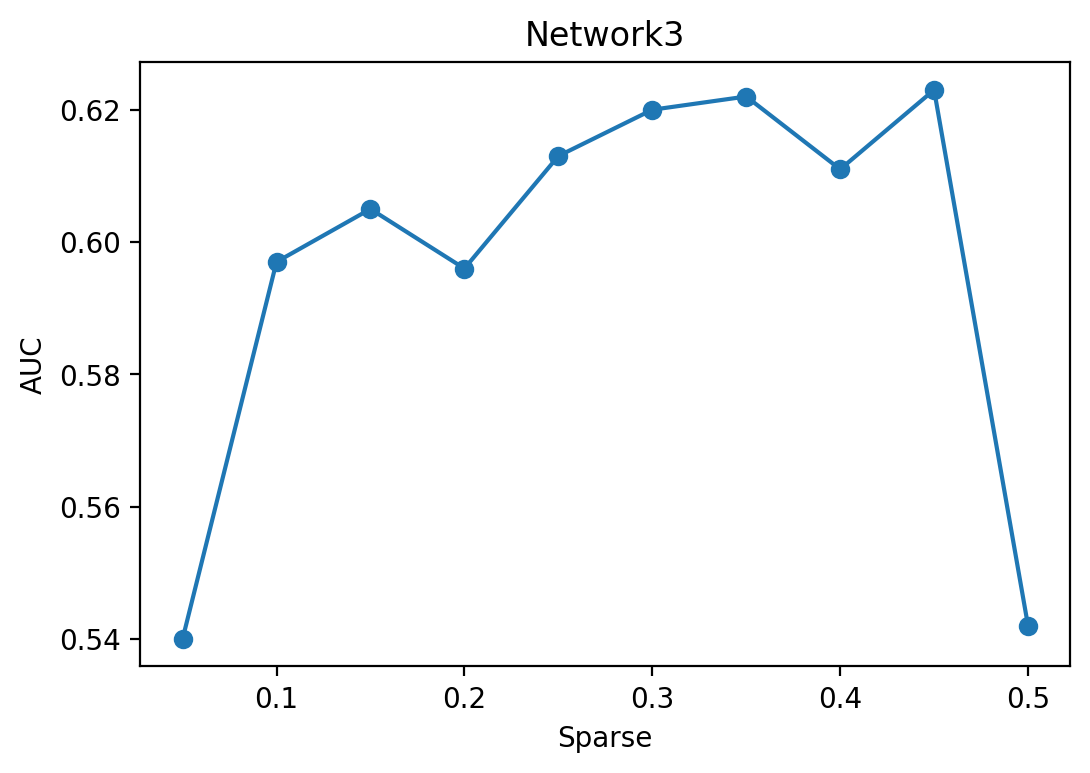}
}
\quad
\subfigure[Network4]{
\includegraphics[scale = 0.4]{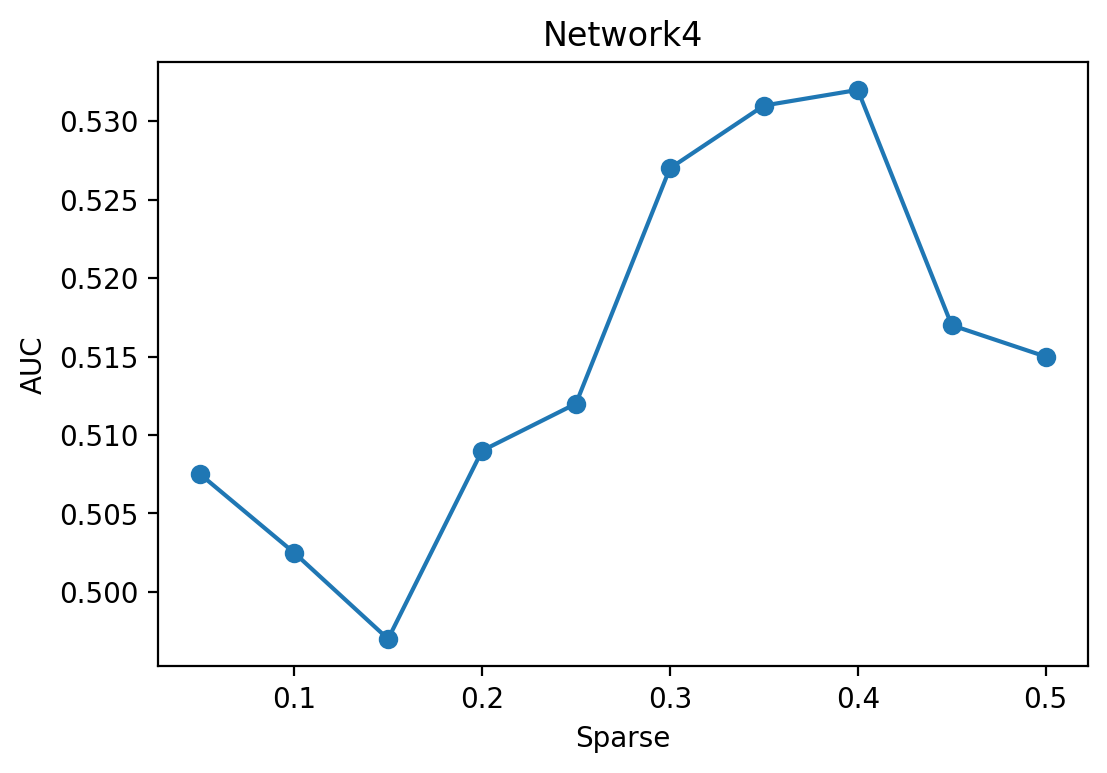}
}
\caption{AUC of four networks with different sparsity}
\end{figure}\par

Considering that it takes more time to reason a matrix containing more non-zero elements, we finally choose the optimal sparsity as 0.55, 0.2, 0.35, 0.35. As the sparsity rises, the time consumed by inference is shown in Figure 6.\par

\begin{figure}[H]
    \centering
\begin{center}
    \includegraphics[scale = 0.4]{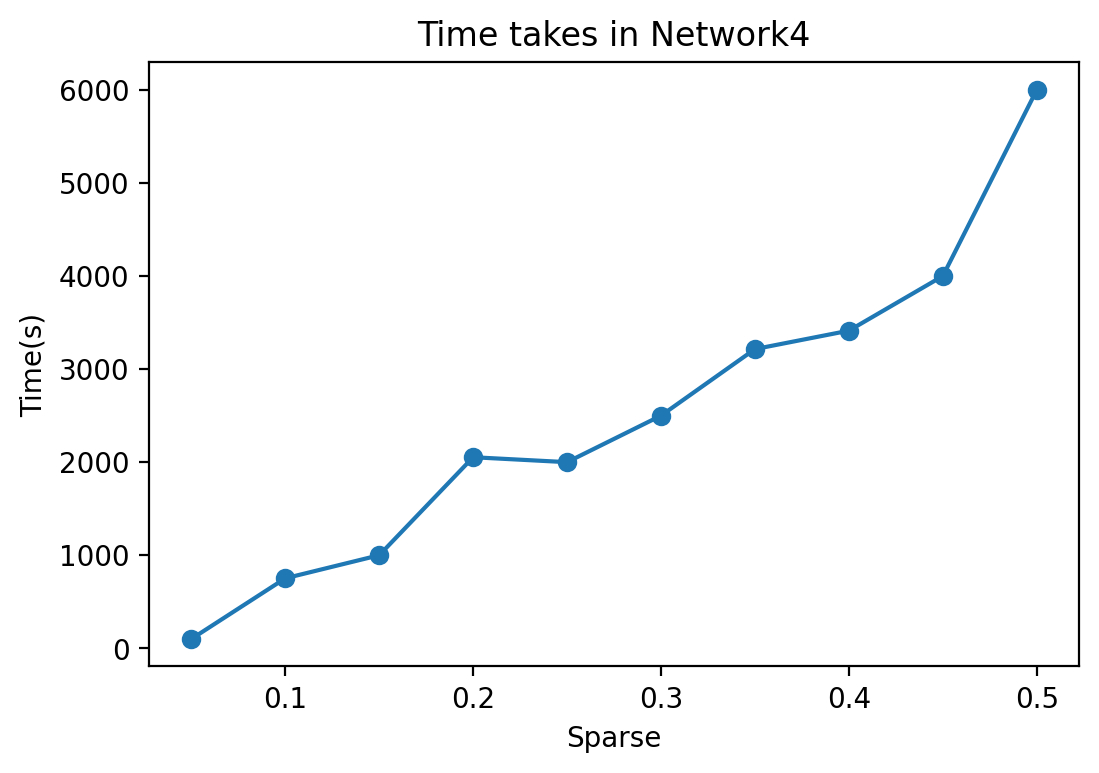}
\end{center}
    \caption{The relationship between time consumption and sparsity}
\end{figure}\par

Meanwhile, the threshold $a$ of the voting algorithm also affects the sparsity of the result. Therefore, we use the same method as above for different values of $a$ and obtain the following results.

\begin{figure}[H]
    \centering
\begin{center}
    \includegraphics[scale = 0.4]{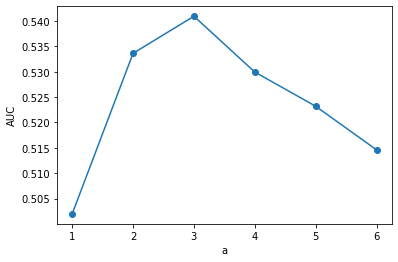}
\end{center}
    \caption{Parameters of the voting algorithm}
\end{figure}\par

The results show that the algorithm has the highest AUC when $a=3$. This means that when more than three of the six independent algorithms give positive judgments, we should assume that there is indeed a moderating relationship here.\par

It is worth noting that the above parameter selection method is applicable to the greedy class of optimization algorithms. For the least squares class of algorithms the selection of the penalty strength coefficient of the regular term is determined using the cross-validation method.

\subsection{Results for the DREAM5 datasets}
Based on the M1 Pro Apple silicon platform, we used the above method to predict the four modulation networks and obtained the following test scores and time consumption:
\begin{table}[H]
\begin{center}

\begin{tabular}{@{}ccccc@{}}
\toprule
                 & \textbf{Network1} & \textbf{Network2} & \textbf{Network3} & \textbf{Network4} \\ \midrule
\textbf{Lasso} & 0.7169            & 0.5554            & 0.5601            & 0.5295            \\
\textbf{OMP}   & 0.6743            & 0.5322            & 0.5806            & 0.5117            \\
\textbf{ISTA}    & 0.6888           & 0.5338            & 0.6177          & 0.5211            \\
\textbf{CoSaMP}    & 0.6115            & 0.6549            & 0.5735            & 0.5117            \\
\textbf{SP}     & 0.6090            & 0.6669            & {\color{red}0.6224}     & 0.5325            \\
\textbf{IHT}     & 0.5013            & 0.7570           & 0.5027            & 0.5049            \\
\textbf{Voting}      & {\color{red} 0.7352}            & {\color{red} 0.7698}            & 0.5962            & {\color{red} 0.5409}            \\ \bottomrule
\end{tabular}
\caption{AUC result}
\end{center}
\end{table}\par

\begin{table}[H]
\begin{center}

\begin{tabular}{@{}lllll@{}}
\toprule
\textbf{}    & \textbf{Network1} & \textbf{Network2}                                     & \textbf{Network3} & \textbf{Network4} \\ \midrule
\textbf{AUC} & 0.809             &0.5845 & 0.65              & 0.528            \\ \bottomrule
\end{tabular}
\caption{result from DREAM5}
\end{center}
\end{table}\par

The results show that the accuracy of our algorithm exceeds the official DREAM5 ranking for Network2, Network3, and Network4 inference.The voting algorithm performs well in the inference of all four networks and is the best performing algorithm in Network1, Network2, and Network4. Except for the prediction for Network2, the IHT algorithm performs poorly with an AUC around 0.5. To our surprise, the traditional Lasso and OMP algorithms, although not the best performers, always give a satisfactory prediction.

\begin{figure}[H]
\centering
\subfigure[Network1]{
\includegraphics[scale = 0.4]{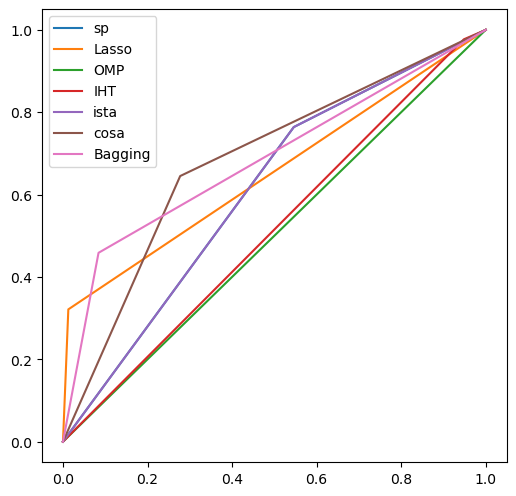}
}
\quad
\subfigure[Network2]{
\includegraphics[scale = 0.4]{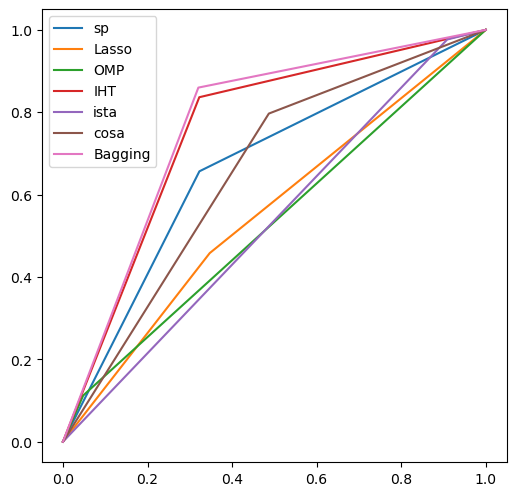}
}
\quad
\subfigure[Network3]{
\includegraphics[scale = 0.4]{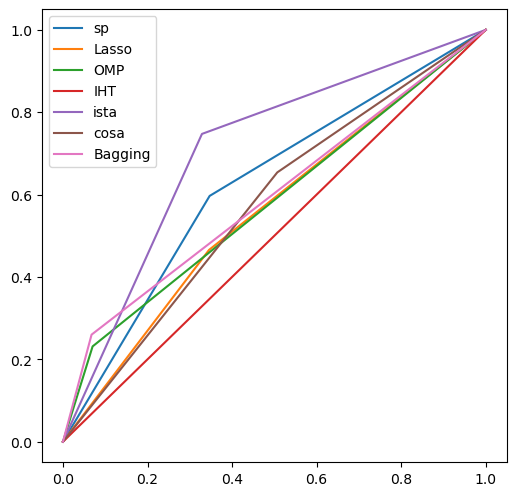}
}
\quad
\subfigure[Network4]{
\includegraphics[scale = 0.4]{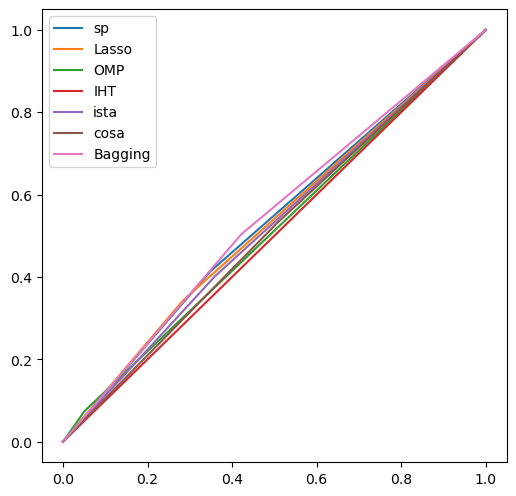}
}
\caption{AUC of four networks with different algorithms}
\end{figure}\par

However, except for the voting algorithm, the performance of the other independent algorithms varies greatly across the data sets. For example, the coordinate descent-based Lasso algorithm performs well in Network1, while it is in the bottom position in the rest of the datasets. The best performer in Network2 was the IHT algorithm, ISTA achieved the highest score in Network3, and the best algorithm in Network4 was the SP algorithm. \par

In terms of time complexity, shown as table 6, the OMP algorithm always takes the shortest time. Among the independent algorithms, CosaMP takes the longest time, even close to 24 hours in the inference for Network3 and Network4.Because the voting algorithm is developed on top of the rest of the independent algorithms, it requires the longest inference time.

\begin{table}[H]
\begin{center}

\begin{tabular}{@{}ccccc@{}}
\toprule
                      & \textbf{Network1} & \textbf{Network2} & \textbf{Network3} & \textbf{Network4} \\ \midrule
\textbf{Lassotime} & 25m24s             & 37m35s           & 53m36s            & 1h21m6s           \\
\textbf{OMPtime}      & 15m06s             & 13m38s             & 31m43s             & 30m41s             \\
\textbf{ISTAtime}     & 5h48m13s             & 1h14m33s             & 8h2m42s             & 7h3m31s             \\
\textbf{CoSaMPtime}     & 4h22m56s             & 42m28s             & 23h19m43s            & 20h22m18s            \\
\textbf{SPtime}      & 3h17m44s             & 43m45s             & 8h4m26s             & 7h4m33s             \\
\textbf{IHTtime}       & 2h21m56s            & 33m19s             & 15h30m46s          & 13h1m53s           \\
\textbf{Votingtime}       & 15h53m05s            & 3h45m15s             & 56h28m25s          & 49h54m32s           \\ \bottomrule
\end{tabular}
\caption{Time consumption of the algorithms}
\end{center}
\end{table}\par

Then, we calculated the average performance and average elapsed time of the four networks and obtained the following figure.

\begin{figure}[H]
    \centering
\begin{center}
    \includegraphics[scale = 0.35]{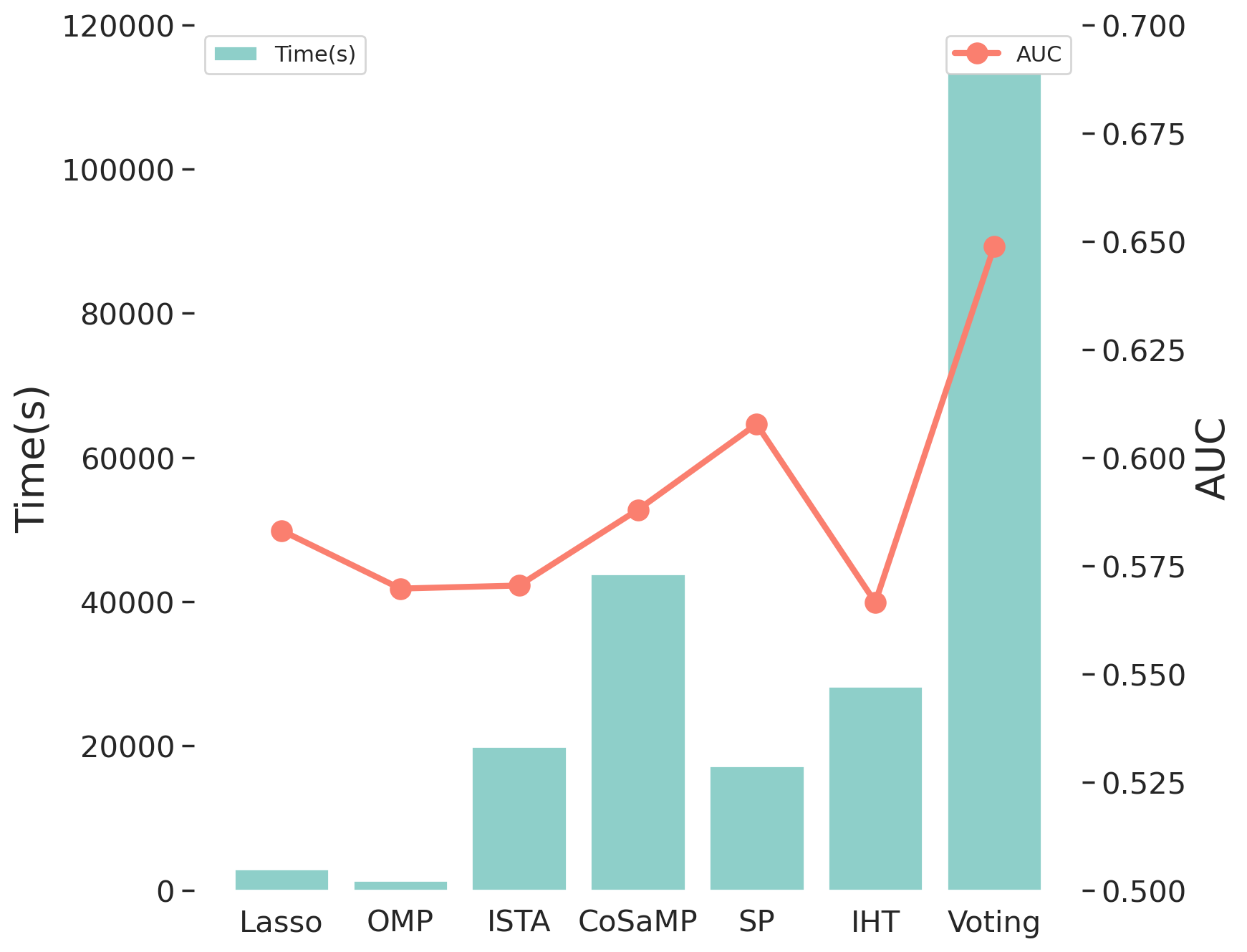}
\end{center}
    \caption{time \& AUC}
\end{figure}\par

From that result we can find that the cost of good performance of the voting algorithm is much higher than the time consumption of the independent algorithm. Among the independent algorithms, SP clearly has the best performance, which takes little time but has the highest accuracy rate.

\subsection{Sensitivity analysis with simulated data}
In the above experiments, we found that for the same algorithm, the accuracy may be related to factors such as the number of transcription factors and the number of samples. Therefore, we used the data generation function of Sparse Tools Box to generate data of different sizes for our experiments. First, we tested the change of the prediction accuracy when the number of transcription factors changed from 10 to 210, controlling other variables constant:

\begin{figure}[H]
    \centering
\begin{center}
    \includegraphics[scale = 0.5]{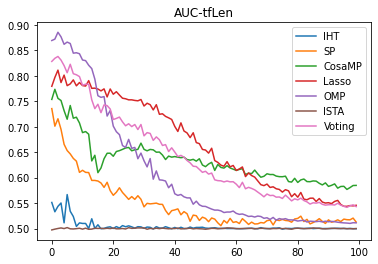}
\end{center}
    \caption{AUC-tfLens}
\end{figure}\par

We can find that the accuracy of all types of algorithms decreases as the number of transcription factors rises. This is due to the fact that the rising number of transcription factors implies the expansion of network size, and the more the system of linear equations to be solved tends to be pathological at a certain number of samples. The difficulty of calculating the exact solution increases greatly at this point.In this set of experiments, the SP algorithm and the OMP algorithm were more influenced by the number of transcription factors. There is a large decrease in accuracy right before the number rises. The CosaMP algorithm, as an improved version of the OMP algorithm, has a good performance, and the accuracy is more stable, and the best result is obtained when the number of transcription factors reaches the maximum.\par
Next, the effect of sample size on accuracy is tested. We set the sample size to vary from 40 to 300 and setting parameters $tfLen = 10$, $geneNums = 200$ and $noise = 0.1$. 

\begin{figure}[H]
    \centering
\begin{center}
    \includegraphics[scale = 0.5]{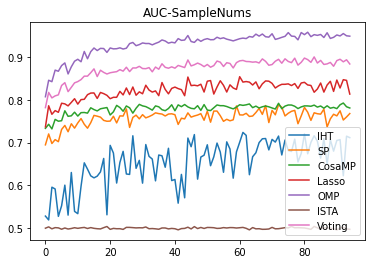}
\end{center}
    \caption{AUC-sampleNums}
\end{figure}\par

The accuracy of each algorithm increases significantly as the number of samples increases. Considering from the perspective of the system of linear equations, as the number of samples rises, the pathological nature of the system of linear equations subsequently decreases and exact solutions are more easily obtained. On the other hand, from the machine learning point of view, the rising number of samples corresponds to providing more independent and identically distributed data for the algorithms to learn. In the ideal case of no noise, as long as these data obey the same distribution, the accuracy of the final learning should also be monotonically undiminished.\par

Finally we consider noise. We let the mean value of the noise vary between 0 and 1, setting the other parameters $tfLen = 10$, $geneNums = 200$ and $sampleNums = 40$.The resulting accuracy variation curves are as follows.

\begin{figure}[H]
    \centering
\begin{center}
    \includegraphics[scale = 0.5]{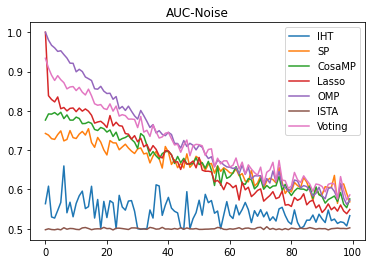}
\end{center}
    \caption{AUC-noise}
\end{figure}\par
As can be seen from the figure, the traditional OMP and Lasso algorithms are more affected by noise, while the CosaMP and SP algorithms and the voting algorithm perform better.\par
Meanwhile, it can be found that OMP, Lasso, CoSaMP, and SP algorithms are not sensitive to hyperparameter response, and the curves are smoother in the above tests. The IHT algorithm, on the other hand, shows a fluctuating curve, which proves that it is sensitive to the response to the value of the hyperparameters.

\subsection{Stress Test}
To test the stability of the package, we will repeat a large number of experiments to record parameters such as average consumption time, variance and mean of accuracy. We tested the data generation-network inference in the low-noise case and the high-noise case, respectively, and the parameters for both cases were chosen as shown in the following:

\begin{table}[H]
\begin{center}

\begin{tabular}{@{}ccccc@{}}
\toprule
              & \textbf{tfLen} & \textbf{sampleNums} & \textbf{geneNums} & \textbf{variance of noise} \\ \midrule
\textbf{Zero} & 10             & 40                  & 200               & 0                      \\
\textbf{Low}  & 10             & 40                  & 200               & 0.15                   \\
\textbf{High} & 10             & 40                  & 200               & 1                      \\ \bottomrule
\end{tabular}
\caption{Pressure test parameter}
\end{center}
\end{table}\par

First we tested the stability of the algorithm in the low-noise and high-noise cases, with the following results.

\begin{figure}[H]
\centering
\subfigure[Low noise]{
\includegraphics[scale = 0.4]{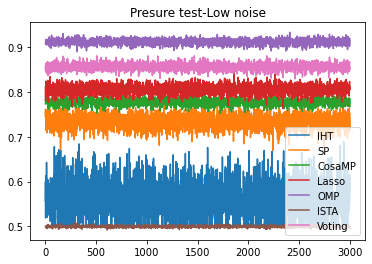}
}
\quad
\subfigure[High noise]{
\includegraphics[scale = 0.4]{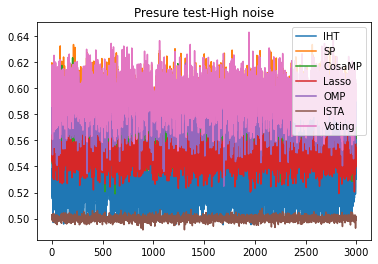}
}
\caption{Stress test in low noise and high noise}
\end{figure}\par

\begin{table}[H]
\begin{center}

\begin{tabular}{ccccc}
\hline
\textbf{}       & \multicolumn{2}{c}{\textbf{Low-noise}}      & \multicolumn{2}{c}{\textbf{High-noise}}     \\
                & \textbf{Std.}        & \textbf{mean-AUC}    & \textbf{Std.}        & \textbf{mean-AUC}    \\ \hline
\textbf{IHT}    & 0.03630         & 0.5596         & 0.01718          & 0.5261         \\
\textbf{SP}     & 0.01478          & 0.7269          & 0.01435          & 0.5892         \\
\textbf{CosaMP} & 0.00759          & 0.7630          & 0.01466          & 0.5728          \\
\textbf{Lasso}  & 0.01014           & 0.7868          & {\color{red} 0.009006} & 0.5481        \\
\textbf{OMP}    & {\color{red} 0.007466} & {\color{red} 0.8758} & 0.01216          & 0.5775         \\
\textbf{ISTA}   & 0.001853         & 0.4999         & 0.001840        & 0.4999        \\
\textbf{Voting} & 0.008284          & 0.8320          & 0.01373          & {\color{red} 0.5934} \\ \hline
\end{tabular}
\caption{Accuracy mean and standard deviation in two cases}
\end{center}
\end{table}\par

From the experimental results, we found that the ISTA algorithm performed poorly in repeated experiments because it relied too much on manual tuning of the parameters, and its score tended to be a random classifier. Therefore, we compare several algorithms other than ISTA, and we find that the coordinate descent-based Lasso algorithm and OMP algorithm have higher stability and accuracy in the case of low noise. For the high noise case, we can find that the mean of accuracy of each algorithm decreases and the standard deviation increases. In this case, our voting algorithm can take advantage of the strengths of all the algorithms and achieve the best prediction results. As we will mention next in the discussion, realistic gene regulatory networks contain a lot of noise. Therefore, this is the reason why we believe that the voting algorithm can solve the problem very well. \par
We also tested the noise-free case, in which the Lasso algorithm with the OMP algorithm was able to reconstruct the gene regulatory network nearly perfectly. 

\begin{figure}

\begin{center}
    \includegraphics[scale = 0.6]{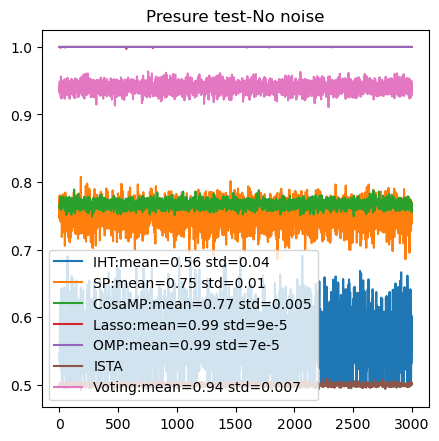}
\end{center}
\caption{No noise case}
\end{figure}

\section{Discussion}
In this paper, we start from the DREAM5 dataset and reason about the four networks it contains using various independent methods with a bagged voting algorithm. Through experiments, we find that there does not exist a general algorithm that can make the reasoning for each network have the best results. The voting algorithm we developed can perform the task better but at the cost of higher time complexity. For the independent algorithms, we can find by analyzing the results that the traditional Lasso regression and OMP algorithms can achieve better results with shorter time consumption for smaller and less sparse networks. However, it was also found that for larger-scale network inference (Network3 \& Network4), improved algorithms such as ISTA and SP were able to provide better prediction accuracy. For the differences arising between traditional and improved algorithms, subsequent experiments based on simulated data offer one possibility: noise. algorithms such as OMP have good convergence for noise-free data, and in some cases, even exact reductions can be obtained. However, as the noise rises, their accuracy decreases very quickly.\par

In a real scenario, we believe that a voting algorithm that takes into account both high and low noise environments is a good choice at the application level since there is no Golden Standard for us to compare scores against.\par

However, when we look at the final obtained prediction network, we also find some interesting facts.The predictor is biased to give a less sparse solution as a way to cover as many of the regulatory relationships proven in Golden Standard as possible. This increases the resulting TPR (the percentage of all samples that are actually positive that are correctly judged as positive) and also increases the FPR (the percentage of all samples that are actually negative that are incorrectly judged as positive).

\begin{figure}[H]
    \centering
\begin{center}
    \includegraphics[scale = 0.15]{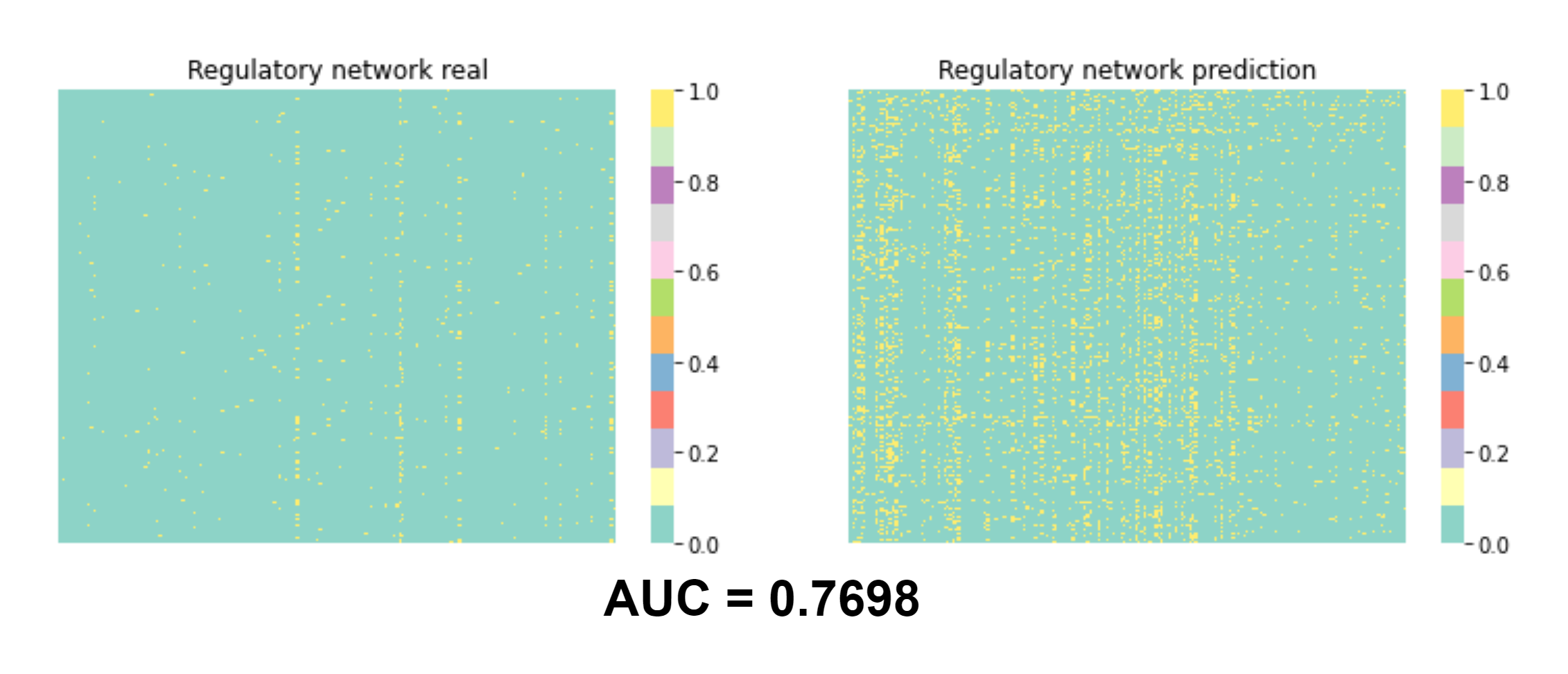}
\end{center}
    \caption{Result GRN of network2}
\end{figure}\par
For a community competition, we believe that submission of all loci that are thought to have a possible regulatory relationship should be encouraged. This is because we can still vote the algorithm on the results of all the submissions that participated in the competition. If there is a locus that most contestants believe has a regulatory relationship, Gold Standard does not. Then it would be a good time to modify Gold Standard.

In the second half of the paper, we focus on the sensitivity of the sparse optimization algorithm to noise on this problem. And it is found that different algorithms will have different performances in different noise environments. Many biologists and mathematicians have experimentally demonstrated that cells are by nature biochemical reactors with noise. Both stochasticity inherent in the biochemical process of gene expression (intrinsic noise) and fluctuations in other cellular components (extrinsic noise) contributes substantially to overall variation. Transcription rate, regulatory dynamics, and genetic factors control the amplitude of the noise. These results establish a quantitative foundation for modeling noise in genetic networks and reveal how low intracellular copy numbers of molecules can fundamentally limit the precision of gene regulation\cite{ref22}. Thattai, M(2001) found in his study that the Fano factor(defined as the ratio of the variance to the mean) is about 20 in the steady-state case. The corresponding steady-state mean value is about 500\cite{ref23}. Therefore, the noise variance can be calculated to be about 10000. This is also a good example of why traditional algorithms do not perform well on real genetic datasets because they lack stability and convergence on noisy data.\par

On the other hand, we also need to focus on the difficulty of obtaining samples. Since there are millions of genes in an organism, we necessarily cannot experiment with millions of living individuals. This means that a linear set of equations of gene regulation is always underdetermined. The application of sparse solution problems of underdetermined linear systems of equations to image processing has been studied extensively\cite{ref24}. This is mainly used to reconstruct a clear image from a noise-laden image. In this study, the sparse optimization method was introduced into the gene regulatory network prediction problem based on the characteristic that gene expression data are difficult to obtain, and it was demonstrated experimentally that good results were obtained. However, this paper is mainly based on the existing algorithms for validation and experiments and does not improve the algorithm from the perspective of the similarities and differences between genetic data and image data. The specific algorithm for genetic data needs to be studied in the future.

\section{Future works}
Although good results have been achieved in this study, there is still some things can be improved. First is for the selection of independent algorithms, we chose three classes each of convex optimization methods and greedy algorithms. More sparse optimization algorithms can be added to the voting system in subsequent studies. In addition, we can also add other non-sparse optimization class algorithms and use their prediction results to further correct the results of the voting algorithm.\par
As for the data, the data used in this study are limited to single-cell organisms, and there is a gap for guiding gene regulation in humans. Therefore, we plan to introduce mammalian genetic data in future studies to further enhance the persuasive power and reference value of the algorithm. \par
For the investigation of the overall nature of gene regulatory networks, this study focuses on the sparsity of gene regulatory networks, but in fact there is another unique and important property of gene regulatory networks: group identity. Group properties mainly describe the presence of identical activation or inactivation between a certain pair or some genes. This property suggests that real regulatory relationships exist not only between genes, but also between groups of genes. A sparse optimization algorithm that incorporates group properties is called a group sparse optimization algorithm(GSO)\cite{ref33}. It can further improve the prediction accuracy and speed up the inference time, in addition to providing a better explanation for the principles of gene regulation.

\end{document}